\documentclass{aa}

\usepackage{graphics,times}

\newcommand{\kms}{\hbox{km\,s$^{-1}$}}
\newcommand{\NH}{\hbox{[N\,{\sc ii}]$\lambda$6583\AA/H$_{\alpha}$}}
\newcommand{\sk}{Sk$-69^\circ279$}

\begin{document}

\msnr{}

\title{Outflow from and asymmetries in the nebula\\
around the LBV candidate Sk-69$\bf^\circ$279}

\titlerunning{The nebula around the LBV candidate \sk}

\author {Kerstin Weis \inst{1,2,}\thanks{Visiting Astronomer, Cerro
Tololo Inter-American Observatory, National Optical Astronomy
Observatories, operated by the Association of Universities for
Research in Astronomy, Inc., under contract with the National
Science Foundation.} \and Wolfgang J.\ Duschl \inst{1,2,}$^\star$
} \offprints{K.\ Weis, Bonn, Germany,
\email{kweis@mpifr-bonn.mpg.de}} \mail{W.J. Duschl, Heidelberg,
Germany}

\authorrunning{K.\ Weis \&\ W.J.\ Duschl}

\institute{Institut f\"ur Theoretische Astrophysik,
Tiergartenstr. 15, 69121 Heidelberg, Germany \and
Max-Planck-Institut f\"ur Radioastronomie, Auf dem H\"ugel 69,
53121 Bonn Germany }

\date{Received / Accepted}

\abstract{We present and discuss new long-slit Echelle spectra of
the LMC LBV candidate \sk\ and put them in context with previous
images and spectra. While at first glance a simple spherically
expanding symmetric shell, we find a considerably more complex
morphology and kinematics. The spectra indicate that
morphologically identified deviations from sphericity are outflows
of faster material out of the main body of \sk. The morphological
as well as the kinematic similarity with other LBV nebulae makes
it likely that \sk\ is an LBV candidate, indeed, and poses the
question in how far outflows out of expanding LBV nebulae are a
general property of such nebulae---at least during some phases of
their evolutions.
}

\maketitle

\section{Introduction}

\begin{figure}
\begin{center}
{\resizebox{\hsize}{!}{\includegraphics{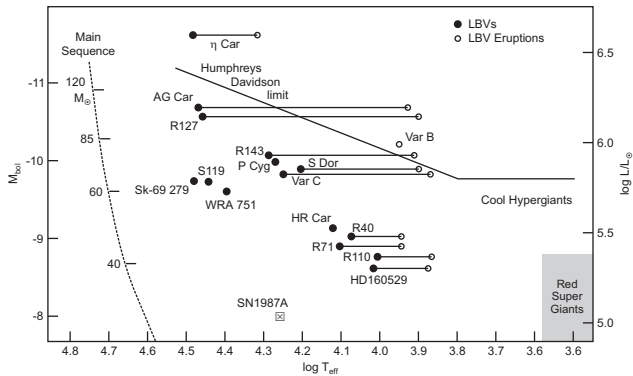}}}
\end{center}
\caption{HRD with the position of known LBVs and LBV candidates.
The position of \sk\ is indicated and similar to that of S119 or
WRA 751, two LBV candidates. For reference the main sequence, the
regime of the hypergiants and red supergiants and the position of
the SN1987A precursor are plotted in addition. \sk\ was positioned
according to the temperature and luminosity determined by Thompson
et al.\ (1982).} \label{fig:hrdlbv}
\end{figure}

Stars are known up to masses around 100\,M$_{\sun}$ with main
sequence luminosities of $10^{5-6}$\,L$_{\sun}$. As O stars they
are located in the upper left part of the {\it Hertzsprung-Russell
Diagram} (HRD). In their later evolution, they leave the
main-sequence at almost constant luminosity towards redder
spectral types, i.e, they quickly cool and become supergiants.
However, instead of completing their evolution towards the red,
the most massive ones among them enter a phase of very high mass
loss (up to 10$^{-4}$\,M$_{\sun}$yr$^{-1}$) and reverse the
direction of their evolution, i.e., the stars become hotter again
(e.g., Schaller 1992; Langer et al. 1994; Schaerer 1996a,b). These
stars are known as {\it Luminous Blue Variables\/} (LBVs). The
location of the turning points of LBVs in the HRD is called the
{\it Humphreys-Davidson limit} (HD-limit; Humphreys \& Davidson
1979, 1994, Langer et al.\ 1994). It is a function of the stellar
luminosity. The position of most LBVs and LBV candidates known is
illustrated in Fig\footnote{ This figure is nearly identical to
Fig.\ 9 from Humphreys \& Davidson (1994), and was copied with the
permission of the authors.}.\ \ref{fig:hrdlbv}, where a solid line
marks the empirical HD-limit. Since LBVs show also spectral
variability both the hottest and coolest temperature (e.g., their
position in the non-eruptive and eruptive states) are indicated
with filled and open dots. The regime of the hypergiants, red
supergiants and the location of the main-sequence is shown for
reference, as is the position of the precursor of SN1987A.
Presently, only a few LBVs are known (roughly 40, including
candidate objects, see, e.g., Humphreys \& Davidson 1994), of
which 9 are in our Galaxy and 10 in the LMC.

The strong stellar winds and possible so-called giant eruptions,
lead to the formation of nebulae around LBVs, the {\it LBV
nebulae\/} (e.g., Nota et al.\ 1995, Weis 2001). These LBV
nebulae are typically up to 2\,pc in diameter. Because of this
small size, they can be studied only in our galaxy and---with the
high-resolution of the {\it Hubble Space Telescope\/} (HST)---in a
few neighboring galaxies like, for instance, the {\it Large
Magellanic Cloud\/} (LMC). For a better understanding of the
evolution of LBVs and especially the formation of LBV nebulae a
good knowledge of the parameters of the nebulae around LBVs is of
great interest.

\section{The LBV candidate Sk-69$^\circ$279 and its nebula}

\sk\ (Sanduleak 1970) is a bright and hot star ($m_{\rm pg}
=12\fm5$) in the LMC in the field of the supergiant shell LMC\,2,
or---more precisely---on the eastern edge of the H\,{\sc ii}
complex N\,160 (see Henize 1956). The star was identified as a
H$_{\alpha}$ emission-line star, by Bohannan \& Epps (1974), and
named BE(74) 619 in this publication. The authors note in addition
to the stellar emission-line a sharp feature seen bluewards of
H$_{\alpha}$, but do not discuss this feature further. A
photometry of \sk\  measured the blue supergiant with $m_{\rm V} =
12\fm79$ and $({\rm B} - {\rm V}) = 0\fm05$ (Isserstedt 1975).
Later---based on prism spectroscopy---spectral types of O $-$ B0
(Rousseau et al. 1978) and---based on grating observations---O9f
(Conti et al. 1986) were quoted. Thompson et al.\ (1982) derived
an effective temperature $T_{\rm eff} = 30\,300\,$K, an absolute
bolometric magnitude $M_{\rm bol}=-9\fm72$, $M_{\rm V}=-6\fm77$
and a radius of R$=28$\,R$_{\sun}$. Various analysis especially of
UV data (IUE) and  IR photometry of \sk\ exist (Morgan \& Nandy
1982, Nandy et al.\ 1981, 1984, Gummersbach et al.\ 1995, Misselt
et al. 1999) but were mainly used for the determination of the
interstellar extinction towards the LMC and don't yield further
information concerning the star itself, beside that it is reddened
and that the  stellar wind terminal velocity is of the order of
1000\,\kms (Nandy et al.\ 1984, determined from the Si\,{\sc iv}
line). In a recent analysis of IUE spectra, Smith Neubig \&
Bruhweiler (1999) found that \sk's UV spectra resemble that of
type B0\,{\sc ii}, and note that the star shows weak Si\,{\sc iv},
C\,{\sc iv}, Al\,{\sc iii}, Fe\,{\sc iii} as well as strong
Si\,{\sc ii}/{\sc iii} lines. The most recent photometry
(Schmidt-Kaler et al. 1999) of \sk\ finds values which
are---within the errors---consistent with those of Isserstedt
(1975).

Weis et al.\ (1995; in the following referred to as W95) found
that \sk\ is surrounded by a circumstellar shell which has a
diameter of about 18\arcsec\ or 4.5\,pc (assuming a distance to
the LMC of 50\,kpc). H$_{\alpha}$ images of the shell (Weis et al.
1997, in the following referred to as W97) show a ring-like
structure with some brighter areas within this ring (Knot E, Knot
S and Knot N). Knot E reaches beyond the shell and clearly deforms
the otherwise almost perfectly round ring structure. The shell is
expanding very slowly with an expansion velocity $v_{\rm exp}
\approx 14$\,\kms\ (W97). In addition to an otherwise almost
spherical expansion (see, in particular, the flat expansion
pattern in Fig.\ 4 in W97) kinematic deviations from a simple
expanding shell were found. A feature, which was identified with
the morphologically classified Knot E, moves $\sim 40$\,\kms\
faster than the center of expansion or about 20\,\kms\ faster than
the backside of the shell at the corresponding position. One of
the most unusual properties of the nebula around \sk\ is the high
\NH\ ratio, which reaches a value of 0.7, compared to a \NH\ ratio
of the background H\,{\sc ii} region of only 0.07. In particular
due to the large \NH\ ratio and the star's location in the HRD,
W97 concluded that \sk\ actually is a good candidate for a LBV
with a typical LBV nebula around it. LBV nebulae are characterized
by such large \NH\ ratios due to CNO-processed material.

\begin{figure}
\begin{center}
{\resizebox{\hsize}{!}{\includegraphics{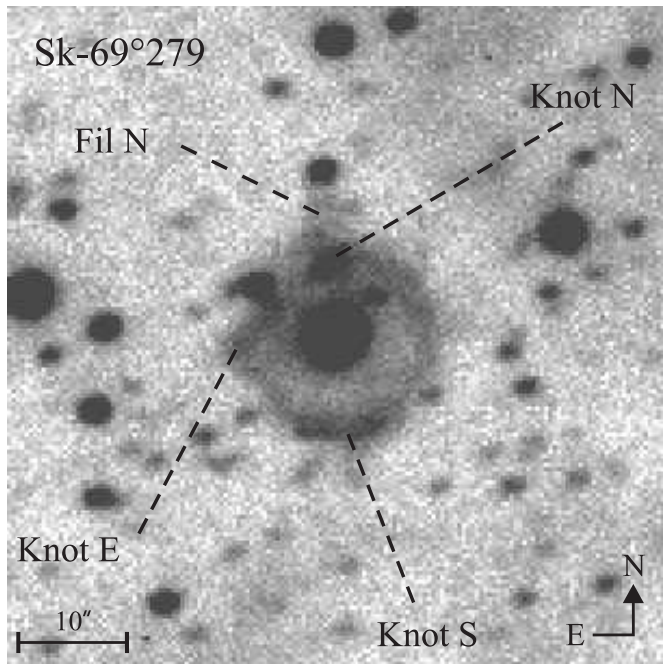}}}
\end{center}
\caption{Image of \sk\ and its nebula, taken with an H$_{\alpha}$
filter and with sub-structure in the nebula marked (see also W97).
The full image is 60\arcsec\,$\times$\,60\arcsec\ large, north is
up and east to the left.} \label{fig:sk69279}
\end{figure}

\begin{figure}
\begin{center}
{\resizebox{\hsize}{!}{\includegraphics{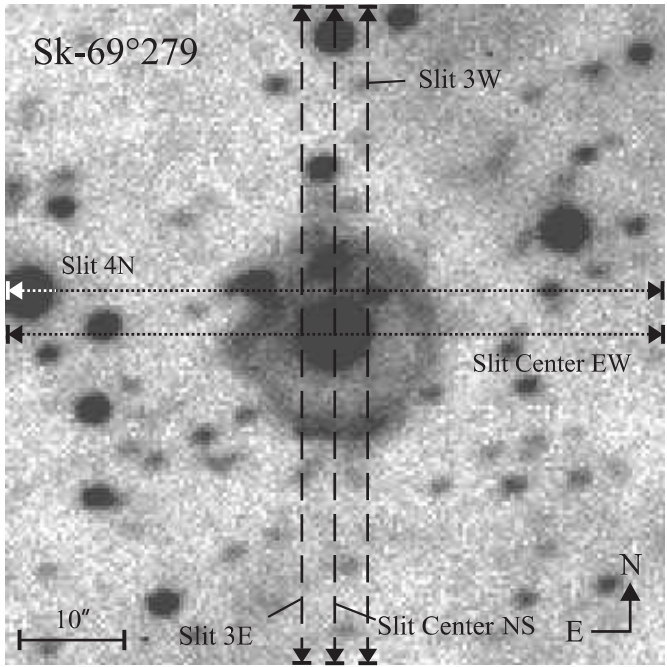}}}
\end{center}
\caption{Here the same images as in figure \ref{fig:sk69279}
is shown of \sk\, now with the orientation of the slits and marked with
the corresponding names.}
\label{fig:nomslit}
\end{figure}

Supporting evidence for \sk\ being a candidate LBV came from the
IUE analysis of Smith Neubig \& Bruhweiler (1999). They detected
in \sk's spectra features which are found in and regarded as
characteristic for the previously classified LBVs and LBV
candidates. In a very detailed analysis about S Dor variables in
the Galaxy and the Magellanic Clouds (van Genderen 2001), stars
which show a certain varibility which is very typical for LBVs,
\sk\  appears among the ex-/dormant S Dor variables\footnote{There
is, however, no history of variability of \sk\ so far, mainly due
to the lack of data (van Genderen, priv. comm.)}. In the same
group as \sk\ van Genderen lists, for instance, S\,119, S\,61 and
He\,3-519---all stars are known as LBV candidates (Humphreys \&
Davidson 1994). Comparing the stellar parameters of \sk\ with
those of other LBVs, the star is only slightly hotter than most of
the other LBVs as seen from Fig.\ \ref{fig:hrdlbv}. \sk's location
is, for example, still very similar to that of WRA\,751 a galactic
LBV candidate or to that of S\,119 (see again Fig.\
\ref{fig:hrdlbv}), another LBV candidate and similar in
temperature to it (see also van Genderen 2001, Tab. 8). Therefore
the star's location in the HRD (see Fig. \ref{fig:hrdlbv}), its
classification as ex-/dormant S Dor variable and its [N\,{\sc ii}]
bright nebula makes it quite likely that \sk\ belongs to the LBV
group. What we do know about the star, is consistent with its LBV
candidate classification. The stars slightly hotter temperature
can probably be attributed to it being a more evolved LBV, as it
was proposed for He 3-519 (Davidson et al.\ 1993).

\section{Observation and data reduction}

\subsection{Imaging with the 0.9\,m CTIO}

For the present investigations, we made use of our already
published image of \sk\ and its nebula (W95, W97). The image was
taken with the 0.9\,m telescope at the {\it Cerro Tololo
Inter-American Observatory\/} (CTIO). A 75\,\AA\ wide H$_{\alpha}$
filter, centered at 6563\,\AA\ was used, which also contains
emission from the [N\,{\sc ii}] lines at $\lambda\,6548$ and
6583\,\AA. The full field of view with this setup was 13\farcm5
$\times$ 13\farcm5 and the scale was about 0\farcs4 pixel$^{-1}$.
The exposure time was 600\,s, the seeing $\sim 1\farcs6$.

\subsection{Long-slit echelle spectroscopy\label{sect:echelle}}

To investigate the kinematics in the \sk\ nebula, new
high-resolution long-slit echelle spectra were taken with the CTIO
4\,m telescope. The setup of the echelle spectrograph was nearly
identical to that used by W97: In both setups the cross-disperser
was replaced by a flat mirror and a post-slit H$_\alpha$ filter
(6563/75\,\AA) was used to select the orders. The spectral region
chosen contained the H$_{\alpha}$ line and the two [N{\sc ii}]
lines at $\lambda\,$6548 and 6583\,\AA. The spatial slit length
was vignetted  to $\sim4^\prime$ and we used the 79\,l\,mm$^{-1}$
echelle grating with a slit-width of 150\,$\mu$m (in contrast to
W97, who used a slit width of 250\,$\mu$m). This leads to an
instrumental FWHM at the H$_\alpha$ line of $\sim
8$\,km\,s$^{-1}$. The data were recorded with the long focus red
camera and a $2048 \times 2048$\,pixel squared CCD, with a pixel
size of 0.08\,\AA\,pixel$^{-1}$ along the dispersion and
0$\farcs$26\,pixel$^{-1}$ on the spatial axis. The spectra were
taken on January 29, 1999. The seeing was $\sim 1$\farcs2 during
the observations. The weather was not photometric. For wavelength
calibration and geometric distortion correction, Thorium-Argon
comparison lamp frames were taken. The telluric lines visible in
the spectra helped to improve the absolute wavelength calibration.
The final accuracy of the wavelength calibration is estimated to
be at least 0.04\,\AA\ or about 2\,\kms.

The data reduction was perfomed using the {\it longslit} package
in IRAF\footnote{IRAF is distributed by the National Optical
Astronomy Observatories which is operated by AURA, Inc. under
cooperative agreement with the NSF.}. For wavelength calibration a
list of thorium and argon lines ($3180 - 9665$\,\AA, as
pre-installed in IRAF and revised in Feb. 1992, Daryl Willmarth,
see IRAF help on linelists) was used, and a Chebychev 6th order
fit. For the correction of the deformation along the spatial axis,
frames with stars present were used. In this case, six stars were
available, located at different positions on the CCD. Here a 2nd
order Chebychev function was fitted. After fitting the calibration
files, this database was used and transformed onto the data. After
correcting for the distortion along the spatial axis and the
wavelength calibration, the final fitting of the data, e.g., the
determination of the velocities, was performed using a gauss
function, and correcting for the continuum emission. The accuracy
to which the peak of the gaussian could be fitted to the data
(that is the central velocity) was estimated fitting all lines
present and comparing the offsets (line fits to each individual
line were always identical within the limits of the fitting
accuracy which is smaller than 0.2\,\kms). The errors of the
radial velocity measurements are therefore estimated to be of the
order of $\pm$ 0.5\,\kms. Since this is much smaller than the size
of the symbols used in all {\it position-velocity} diagrams
($pv$-diagrams), no error bars are visible in these plots.

\begin{figure*}
\begin{center}
{\resizebox{13cm}{!}{\includegraphics{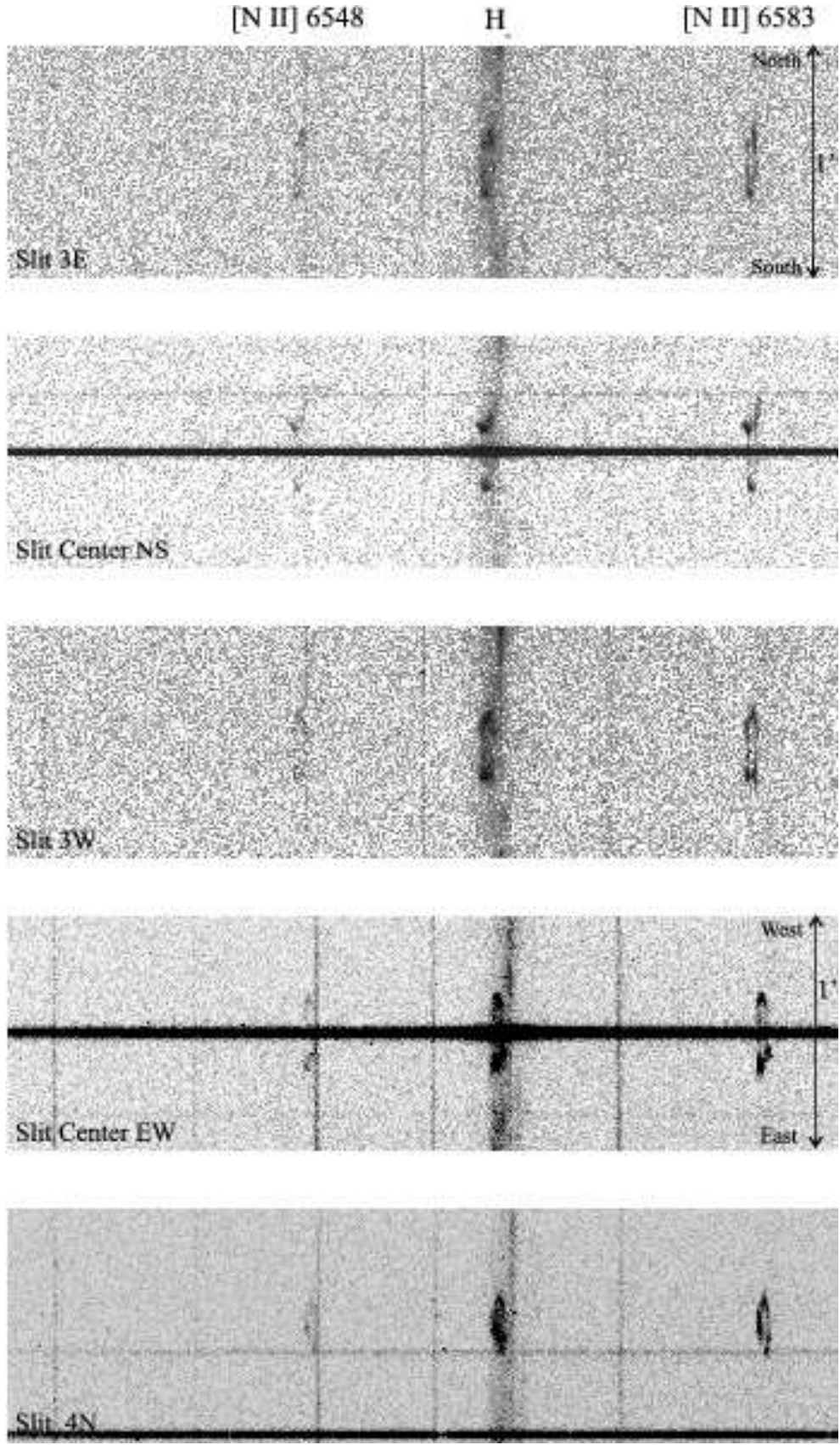}}}
\end{center}
\caption{Echelle spectra for all five slit positions and
orientations. The slit length is $1^\prime$, the slit positions
and orientations are shown in Fig. \ref{fig:nomslit}.}
\label{fig:echellesk69279}
\end{figure*}

To improve on the spatial coverage of our spectral data, we chose
a position angle (PA) of 180$^\circ$, i.e., orthogonal to that of
the spectra by W97. At this PA, three spectra were taken: One
centered on the star ({\it Slit Center NS\/}) and two offset
3\arcsec\ to the east ({\it Slit 3E\/}) and to the west ({\it Slit
3W\/}) of the central star, respectively. The new slit positions
as well as the ones from W97 (dotted lines) are indicated and
labeled in Fig. \ref{fig:nomslit}. For consistency we label W97's
spectra as {\it Slit 4N\/} and {\it Slit Center EW\/}. They were
published originally as Figs. 3 (a) and (b) in W97. The
echellograms of all five spectra are shown in Fig.
\ref{fig:echellesk69279}. The echellograms are 1\arcmin\ in
spatial direction and centered on the star, or---in cases of the
off-centered slits---on the projected position of the star onto
the slit. The spectral range of each echellogram is 65\,\AA\ and
the central position coincides with the rest wavelength of
H$_{\alpha}$.  In Fig. \ref{fig:radialrun} we present the
corresponding $pv$-diagrams of the respective slits. Velocities
are quoted in the heliocentric reference system with the systemic
velocity subtracted, and are determined from the Doppler shifts in
the [N\,{\sc ii}]\ $\lambda\,$6583\,\AA\ line.

\section{The morphology of the nebula around Sk-69$\bf^\circ$279}

A morphological discussion has still to be based on the image
discussed in W97 as no improvement was achieved since. W97 noticed
three deviations in \sk's nebula from a spherical shell,
manifesting themself as extended knots (see also Fig.\
\ref{fig:sk69279}) and named Knots E, S, and N. Knots N and S are
located within the ring structure and are characterized by their
higher surface brightness as compared to their surroundings. In
contrast, Knot E manifests itself as a geometrical deviation from
the ring structure appearing as a triangularly shaped attachment
to the ring. To disentangle these knots from possible
back/foreground stars a continuum subtracted image was made and
shown in W97. While the emission (i.e., the continuum) of all
stars in the field was successfully subtracted, all identified
knots still show up in the continuum free image, giving evidence
that the knots are real brighter emission line regions within the
shell. In addition to these already described features, we
identify in \sk's nebula the Filament N (Fil N, Fig.\
\ref{fig:sk69279}). Fil N extends from Knot N about 7\farcs1
(corresponding to 1.7\,pc) to the north and is---similar to Knot
E---roughly triangular in shape. Fil N has a very low surface
brightness compared to other parts of the nebula and might easily
be mistaken as a part of the background H\,{\sc ii} region.
Because of this, its membership to \sk's nebula can be established
only with the help of a spectral analysis (Sect.\
\ref{sect:spectra}) and thus could not be classified earlier.

\section{The kinematics of the nebula around Sk-69$\bf^\circ$279
\label{sect:spectra}}

Already W97's two spectra of the nebula around \sk\ showed that
the expansion is not perfectly spherical but displays faster
moving small scale features. The three new spectra taken with a
north-south orientation support these earlier results. Fig.\
\ref{fig:echellesk69279} shows the echellograms of all five
spectra. Expansion ellipses are the clearly visible main features
in all spectra.

\begin{figure*}
\begin{center}
{\resizebox{13cm}{!}{\includegraphics{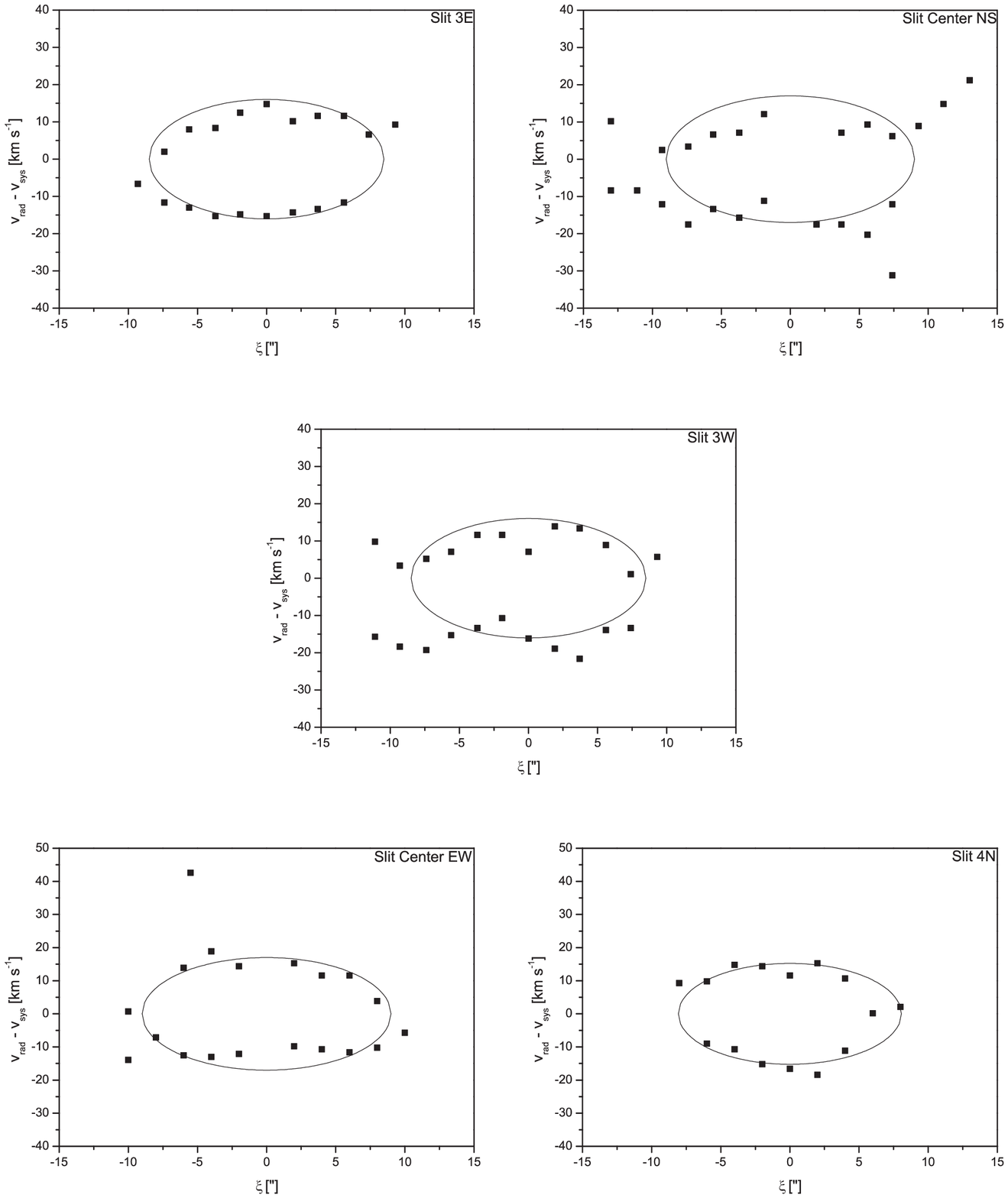}}}
\end{center}
\caption{Position-velocity diagrams for all five spectra of Fig.
\ref{fig:echellesk69279}. The lines give the model results
according to Eq. \ref{eq:pv-model}, while the squares give the
corresponding measurements for a systemic velocity of
230\,km\,s$^{-1}$. The coordinate $\xi$ is measured along the slit
with the projected position of the star at the origin. Negative
offsets are to the south (Slits Center NS, 3W and 3E) or to the
east (Slits Center EW, 4N).} \label{fig:radialrun}
\end{figure*}

For comparison purposes, we model the expansion as being
spherically symmetric with a velocity $v_{\rm exp}$ and a radius
$R$ of the nebula. We introduce a coordinate system $\left\{ \eta,
\xi \right\}$, where the coordinate $\xi$ is measured along the
slit and has its origin at the star's projected position, and the
coordinate $\eta$ in a perpendicular direction with the same
origin. The observed radial velocity distribution $v_{\rm rad,
rel}^{\rm spher}$ relative to the systemic velocity then is

\begin{equation}
\left| v_{\rm rad, rel}^{\rm spher} (\eta, \xi) \right| = v_{\rm
exp} \left( \frac{R^2 - \eta^2 - \xi^2}{R^2} \right)^{1/2}.
\label{eq:pv-model}
\end{equation}
In Fig. \ref{fig:radialrun} we compare for all five slit positions
and orientations the model velocities with the observed values.
The measured radial velocities were transformed into the central
star's reference system by correcting for a systemic velocity of
230\,km\,s$^{-1}$ (W97). We find that in all cases the observed
run of the radial velocities differs from what a pure spherical
model predicts. These differences are significant given the
accuracy of our spectra.

In addition, a closer comparison of the $pv$-diagrams (Fig.
\ref{fig:echellesk69279}) shows that the maximum expansion
velocities (here defined as the center of the line split) of each
Doppler ellipse lie at different positions in each slit. In Slit
3E the maximum expansion velocity amounts to 15.1\,\kms\ at the
position 0\arcsec, in Slit Center NS it is 14.8\,\kms\ at
5\farcs7, in Slit 3W 17.5\,\kms\ at 3\farcs7, in Slit Center EW
16.0\,\kms\ at $-4\arcsec$, and in Slit 4N 16.9\,\kms\ at
2\arcsec. These off-centered maximum expansion velocities and the
asymmetric shapes of the ellipses are a clear indication that the
expansion is not spherically symmetric. In particular in Slit 3W
(see middle panel in Figs. \ref{fig:echellesk69279} and
\ref{fig:radialrun}) the expansion pattern is deformed and
considerably deviates from an elliptic shape, being smaller at the
center (see waist like deformation at position 0\arcsec) and
getting wider to the north at $\sim 4\arcsec$.

Comparing the positions of the maximum expansion velocity of each
slit we find that the regions with the highest expansion
velocities are located nearly exactly to the north (since the
maximum expansion velocity found in the north-south oriented slits
lie at positive offsets) and to the east of the star. Two of the
spectra (Slit 3E and Slit 4N) have expansion ellipses which are
open on one side, in Slit 3E to the north (upper part) and in Slit
4N to the east (lower part). Here the shell forms no coherent
structure anymore, or only very faint emission with a much lower
surface brightness (below our detection limit) is present.

Beside of this global expansion pattern, several regions are found
in the spectra which move significantly faster or slower than the
expansion ellipse at that point or extend beyond the ellipse. Most
prominent is a feature seen in Slit Center NS (Figs.
\ref{fig:echellesk69279} and \ref{fig:radialrun}). Here to the
north (top) of the slit a long red-shifted emission extends
beginning at the end of the ellipse. This feature is about
10\arcsec\ long, corresponding to 2.5\,pc. The spectrum shows that
with increasing distance the velocity becomes larger and increases
from the edge of the shell at 6.0\,\kms\ by about 15\,\kms\ to
21.0\,\kms\ at our outermost measurement. This faster moving
structure can be identified with the morphologically classifies
Filament N. A corresponding fainter emission can be traced in
Slits 3E and 3W to the north end of the expansion ellipses. This
emission is also a contribution of Filament N, which is partly
intercepted by these slits (see position Slit 3W, Slit 3E in Fig.
\ref{fig:nomslit}). Also at the northern rim (at $\sim 6\arcsec$)
of the ellipse in Slit Center NS appears a structure which shows a
brighter knot-like emission and is blue-shifted with respect to
the expansion ellipse at that position. The highest blue-shifted
emission of the structure is at about $-31.0$\,\kms\ and thus
20\,\kms\ faster than the expanding shell. The higher surface
brightness and the position makes it likely that this blue-shifted
emission results from Knot N. At the south end ($-10$ to
$-15$\arcsec) of Slit Center NS and Slit 3W we detect a component
moving with approximately $+10$\,\kms (Fig.\
\ref{fig:echellesk69279}) which is therefore expanding faster than
the shell at this point. This component can be assigned to the
optically identified structure called Knot S. Finally we note that
Slit Center EW has a highly red-shifted knot at +42\,\kms\ and was
already identified with Knot E in W97.

\section{Discussions and Conclusions}

The images of the nebula around \sk\ as well as the spectra show
several cases of deviations from the morphology and kinematics of
a spherical nebula with symmetric expansion.

Due to the lack of sufficiently resolved images of the nebula, one
can only speculate---with the help of spectra---about the three
dimensional structure of the nebula. While part of its main body
may reasonably well be approximated by a sphere with a symmetric
expansion, the nebula is expanding faster in the north-eastern
hemisphere comparing, for instance the higher expansion velocities
detected in Slits 3W (17.5\,\kms) and Slit 4N (16.9\,\kms) with
those in the nebula's center (typically 15\,\kms, Slit 3E or Slit
Center NS) or the southern part of the nebula which expands even
slower than the center (see $pv$-diagrams in Fig.
\ref{fig:radialrun}). Even though the difference of the expansion
velocities is marginal, it is significantly higher than the
expected errors, and therefore most likely real.

Knot E shows an extension of the nebula of about 3\farcs2
(0.8\,pc) and moves faster than the shell (red-shifted). Here
material seems to surpass the shell and move away from the
observer. Similarly, the material of Knot N approaches us faster
than most of the shell (see blue-shifted extension in Slit Center
NS). In addition to this faster moving material of Knot N, there
gas is streaming out. This out-streaming gas---as traced in the
spectra---corresponds to the morphologically identified Fil N. Fil
N has a length of roughly half the diameter of the shell and moves
up to 15\,\kms\ faster along the line-of-sight. Throughout the
nebula---including Fil N---we find a \NH\ ratio in the range of
$0.65 - 0.7$ and thus further evidence for Fil N being a part of
\sk's nebula.

The spectra also indicate that to the north-east the shell of the
nebula around \sk\ is not closed. Together with the faster moving
Fil N it is therefore most likely that the nebula around \sk\
shows an outflow. The north-eastern part of the nebula has opened
up and material is streaming out evidenced in particular by Fil N.
Both properties, open Doppler ellipses and an increasing velocity
extension, were also found in the nebula around the LMC LBV
candidate S\,119 (Weis et al.\ 2002). The kinematics and
morphology of the nebula around S\,119 are very similar to those
found here. In particular, it is worthwhile noting that this
outflow from \sk's nebula, shows the same linear increase in
velocity as the outflow in S\,119. The out-streaming material
moves the faster the further it is away from the star. Comparing
both objects we conclude that the nebulae of these stars are quite
similar and represent (candidate) LBV nebulae with outflows.

It is not clear whether Knot E is also part of an outflow or if
this feature resembles more an asymmetry in the nebula comparable
to the Caps in WRA\,751 (Weis 2000) or R\,127 (Weis 2002, in
prep). An interpretation as an outflow or as caps is supported by
the faster moving extension in Slit 3W.

The dynamic age of a nebula is defined by $\tau = \eta (r/v_{\rm
exp})$, with $\eta$ depending on whether we model a steady fast
wind swept up bubble ($\eta=1$, Garc{\'\i}a-Segura \& Mac Low
1995), an energy-conserving bubble ($\eta = 0.6$, Weaver et al.\
1977) or a momentum-conserving bubble ($\eta=0.5$, Steigman et
al.\ 1975). For \sk\ this would yield a range for the dynamic age
between $6.3\,10^4$ (for a maximum expansion velocity of
17.5\,\kms\ and a momentum conserving bubble) to $1.5\,10^5$ years
(with the slowest expansion of 14\,\kms\ and steady sweep-up).
Compared to dynamic ages of other LBV nebula (ranging typically
between $0.01-2\,10^4$, see, e.g., Nota et al. 1995) this value is
up to a factor of 10 higher than the average. With the estimated
duration of the LBV phase of about 25\,000 years (Maeder \& Meynet
1987, Humphreys \& Davidson 1994, Bohannan 1997) the dynamic age
would be too high for the nebula being created during the stars
LBV phase. In this context, however, one has to keep in mind that
the duration of the LBV phase is mainly estimated by determining
the ratio of LBVs stars to Wolf-Rayet stars, i.e., a method with
comparatively large inherent errors, due to incompleteness of the
respective samples. In the context of \sk\ beeing a nebula with an
outflow, as proposed here, these age determinations with the help
of the nebula's size and expansion velocity might well be
overestimated. If the nebula was disrupted by the outflow, most
likely the expansion velocity of the nebula decreased since the
pressure dropped. Therefore the dynamic age determination yields a
higher age, as a result from the lower expansion velocity.  The
dynamic age for S\,119 another LBV candidate showing outflow, for
instance, also seems to be slightly higher compared to other LBV
nebulae, reach a  maximum of $\sim 4\,10^4$ years (Weis et al.\
2002).

While from the kinematics alone it is not clear yet clear in each
case which features are part of an outflow and which are cap-like
extension of the shell, the structure of the expansion ellipses
as well as Fil N show that at least these features are due to an
outflow. The similarity with other LBVs and their nebulae makes it likely
that \sk\ is at least a good candidate LBV, indeed, and poses the
question in how far
outflows out of expanding LBV nebulae are a general property of
such nebulae---at least during some phases of their
evolutions---and thus connects directly to the question for their
origin and evolution.

\begin{acknowledgements}

We thank the referee for helpful comments on the manuscript. KW thanks
Dr.\ D.J.\ Bomans (Bochum) for many helpful discussions of the
subject of this paper. Sincere thanks go to Prof.\ R.M.\ Humphreys
for permitting us to reprint and slighlty varify Fig.\ 9 from
Humphreys \& Davidson (1994). The data reduction and analysis was
in part carried out on a workstation provided by the {\it Alfried
Krupp von Bohlen und Halbach-Stiftung\/} to the ITA, Heidelberg.
Their support is gratefully acknowledged.

\end{acknowledgements}

\end{document}